\documentstyle[11pt,newpasp,twoside,epsf]{article}
\def\edcomment#1{\iffalse\marginpar{\raggedright\sl#1\/}\else\relax\fi}
\reversemarginpar

\begin{document}
\title{The Satellite-substructure Connection}
 \author{J.~E.~Taylor (1), A.~Babul (2) \& J.~Silk (1)}
 \affil{(1) Astrophysics, University of Oxford\\ 
Denys Wilkinson Building, Keble Road, Oxford, OX1 3RH, UK\\
         (2) Physics and Astronomy, University of Victoria\\ 
Elliott Building 3800 Finnerty Road, Victoria, BC, V8P 1A1, Canada }
\begin{abstract}
We describe our recent attempts to model substructure in dark matter
halos down to very small masses, using a semi-analytic model of halo 
formation. The results suggest that numerical 
simulations of halo formation may still be missing substructure 
in the central regions of halos due to purely numerical effects. 
If confirmed, this central `overmerging' problem will have important 
consequences for the interpretation of 
lensing measurements of substructure. We also show that the spatial 
distribution 
of subhalos relative to the satellite companions of the Milky Way rules out 
at least one simple model of how dwarf galaxies form in low-mass halos.
\end{abstract}
\section{Introduction}
When numerical simulations of cold dark matter (CDM) 
halo formation reached sufficient 
resolution a few years ago, they revealed that a wealth of dense 
substructure, the 
undigested remains of ancient hierarchical merging, should still 
survive in these systems at the present day (Klypin et al.~1999;
Okamoto \& Habe 1999; Moore et al.~1999). The properties of halo 
substructure will be a sensitive test of the nature of dark matter 
if we can manage to quantify them observationally, for instance in
strong gravitational lens systems (cf.\ talks by P.~Schneider and N.~Dalal 
in these 
proceedings). Halo substructure must also be linked to satellite galaxies, 
such as the dozen dwarf satellites of the Milky Way, although the exact 
nature of the connection remains problematic, given the huge discrepancy
between the number of subhalos predicted by simulations and the number
of luminous satellites observed locally.

Before we can reach firm conclusions on the properties of halo substructure 
or the nature of the satellite-substructure connection, however, we should 
first confirm that current simulations of substructure have converged
to a definite and accurate prediction of its properties. 
For many years, N-body simulations suffered from `overmerging',
the artificial disruption of substructure due to numerical effects.
It is still not clear that all the properties of simulated halo substructure
have converged to level required for applications such as lens modeling.
To test for the possibly of residual overmerging, we will compare purely
numerical results with semi-analytic (SA) predictions, using a recently
developed SA model of halo formation. 

\section{Is Overmerging Over Yet?}
The SA model of Taylor \& Babul (2003, TB03 hereafter) provides an alternative
to direct numerical simulation of halo formation. It combines extended
Press-Schechter merger histories, which predict when subhalos merge, 
with an analytic description of orbital evolution 
and mass loss, which determines how they evolve subsequently. While
the model is approximate in many ways, it also avoids 
some of the strongest biases in the purely numerical studies,
and thus provides an interesting point of comparison.

Fig.\,1a shows the mass spectrum of subhalos in simulations by 
Moore et al.\ (1998, 1999) and Ghigna et al.\ (2000) 
(dotted lines), compared with the SA model of TB03 (thick solid lines, 
with thin solid lines showing the 1-$\sigma$ variance). In the outer
part of the halo, the two techniques agree to within 10--20\%.
In the inner regions, where higher densities and longer 
evolution times would enhance numerical overmerging, the SA model 
predicts up to twice as much substructure at a given mass. In Fig.\,1b,
we compare the spatial distribution of subhalos in simulations 
of increasing resolution (panels from right to left). While the 
amount of central substructure increases slowly with resolution,
there is no obvious sign of convergence -- simulations at even 
higher resolution might find yet more subhalos in the 
central regions, as predicted by the SA model (solid lines).
\begin{figure}
\plottwo{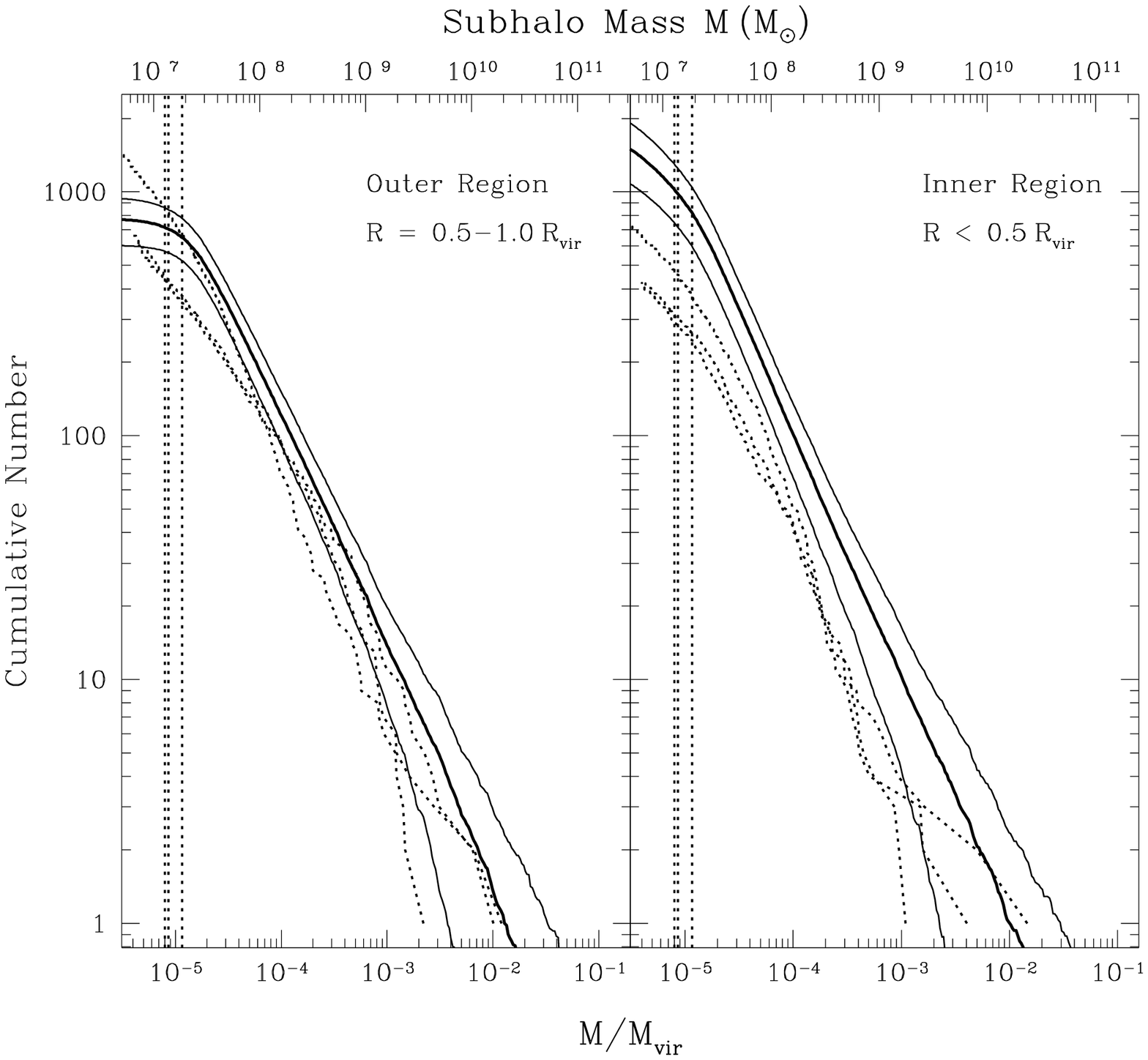}{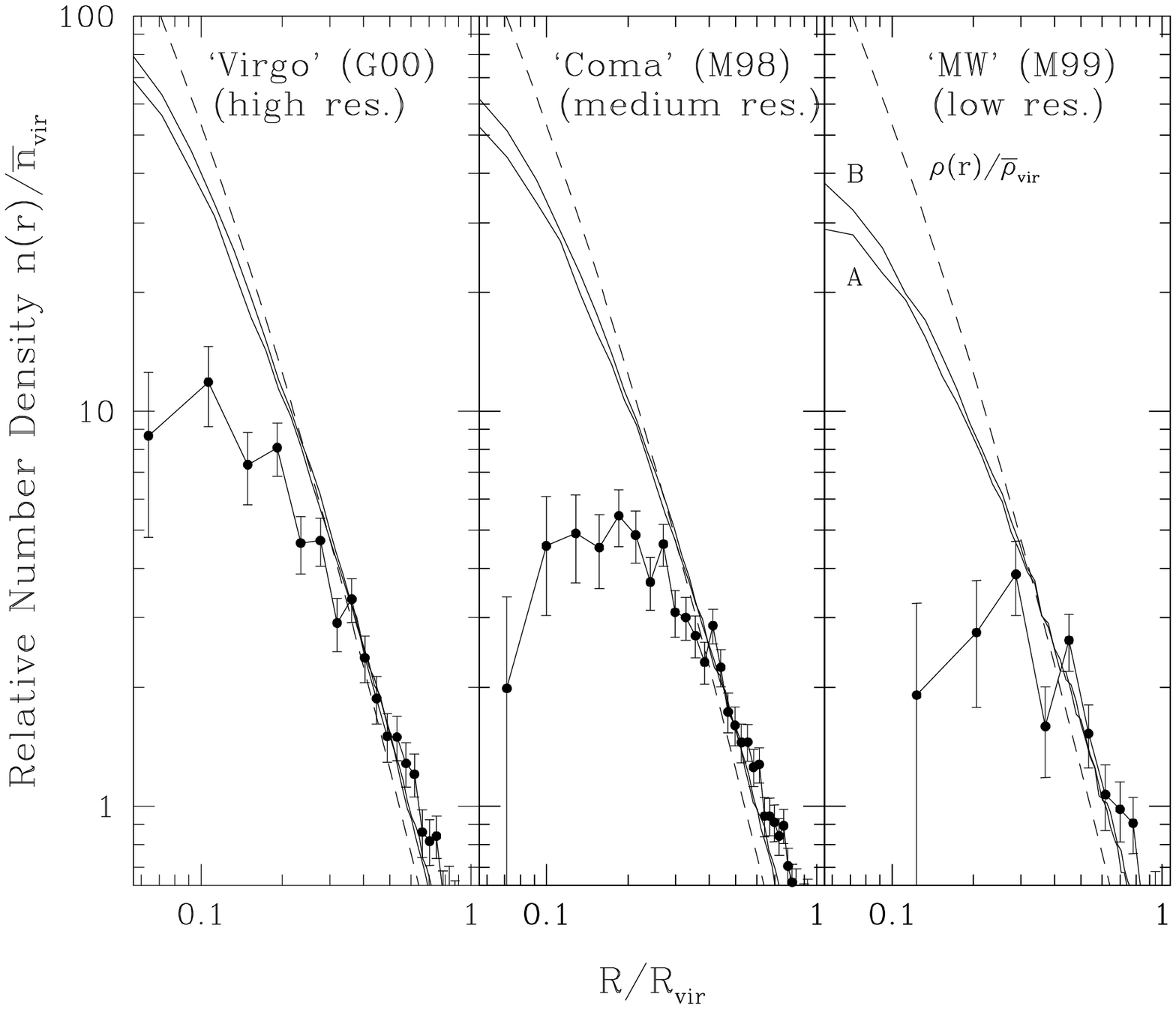}
\caption{a: ({\sl left}) Subhalo mass functions from simulations and from
the SA model of TB03 (dotted and solid lines respectively). b: ({\sl right}) 
The relative number density of subhalos in three different simulations 
(lines with error bars), compared with the SA model (solid lines),
and the background density profile (dashed line).
}
\end{figure}
%
\section{Implications for Lensing Measurements of Substructure}
By examining flux-ratio anomalies in multiply imaged quasars, strong
lensing studies can constrain the amount of substructure along lines of sight
through the center of galactic halos. Robust predictions of the amount of 
central substructure are essential to take full advantage of this work.
Fig.\,2a compares predictions for the number of subhalos and the mass fraction 
in substructure in the highest-resolution simulation and in the SA model. 
The differences discussed previously produce an 
order-of-magnitude increase in the predicted mass fraction in 
the central regions. Interestingly,
the level predicted by the SA model is roughly that inferred from 
observations (e.g.~N.~Dalal, these proceedings). The SA model
also predicts a large scatter in the mass fraction, however. 
Fig.\,2b shows that the
amount of substructure decreases as systems age. Dynamically young
systems (those whose halos have formed recently) have 3--4 times more
mass in substructure then more relaxed systems.
This large scatter implies that sample selection
will be an important factor in lensing studies.
\begin{figure}
\plottwo{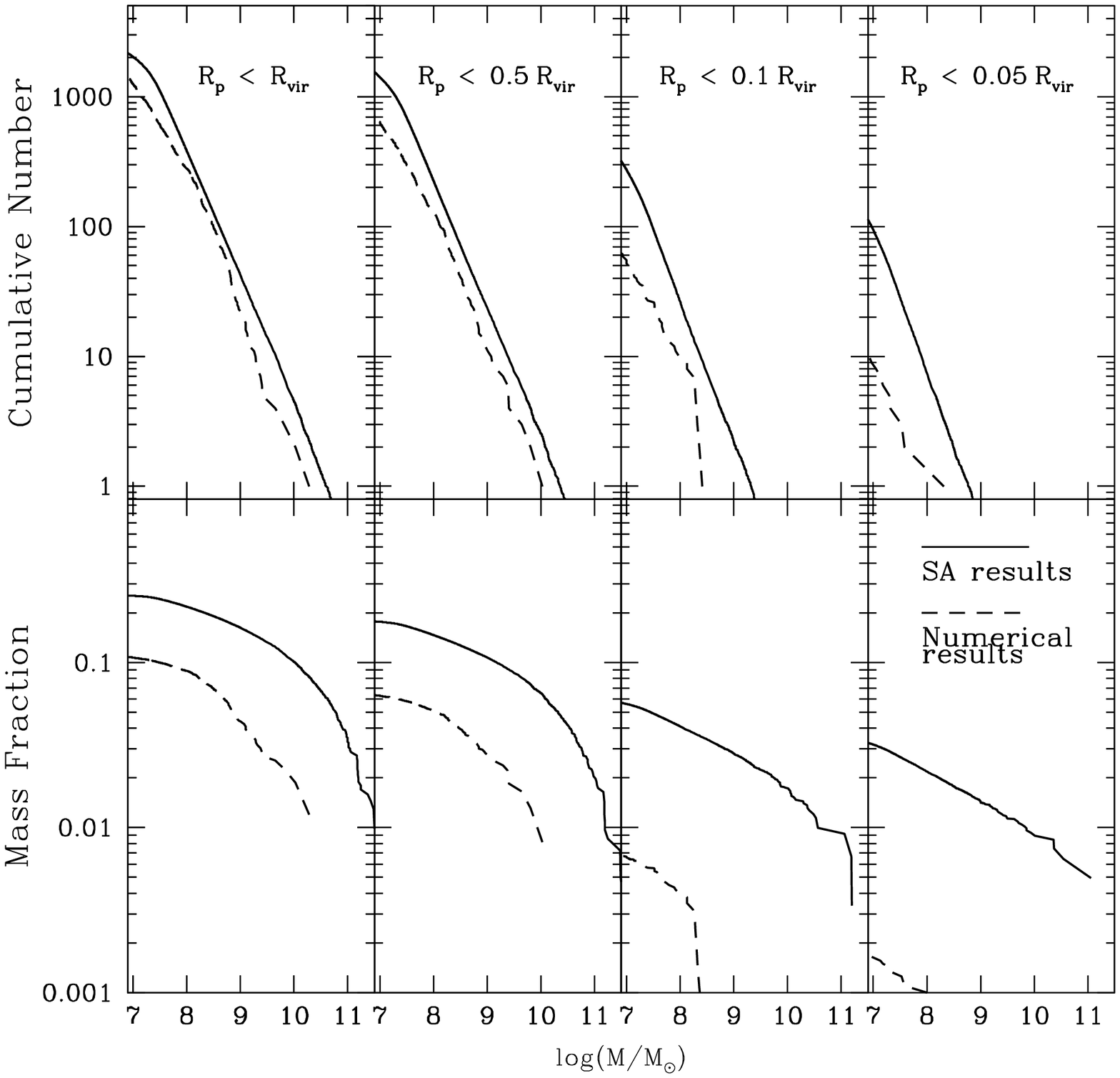}{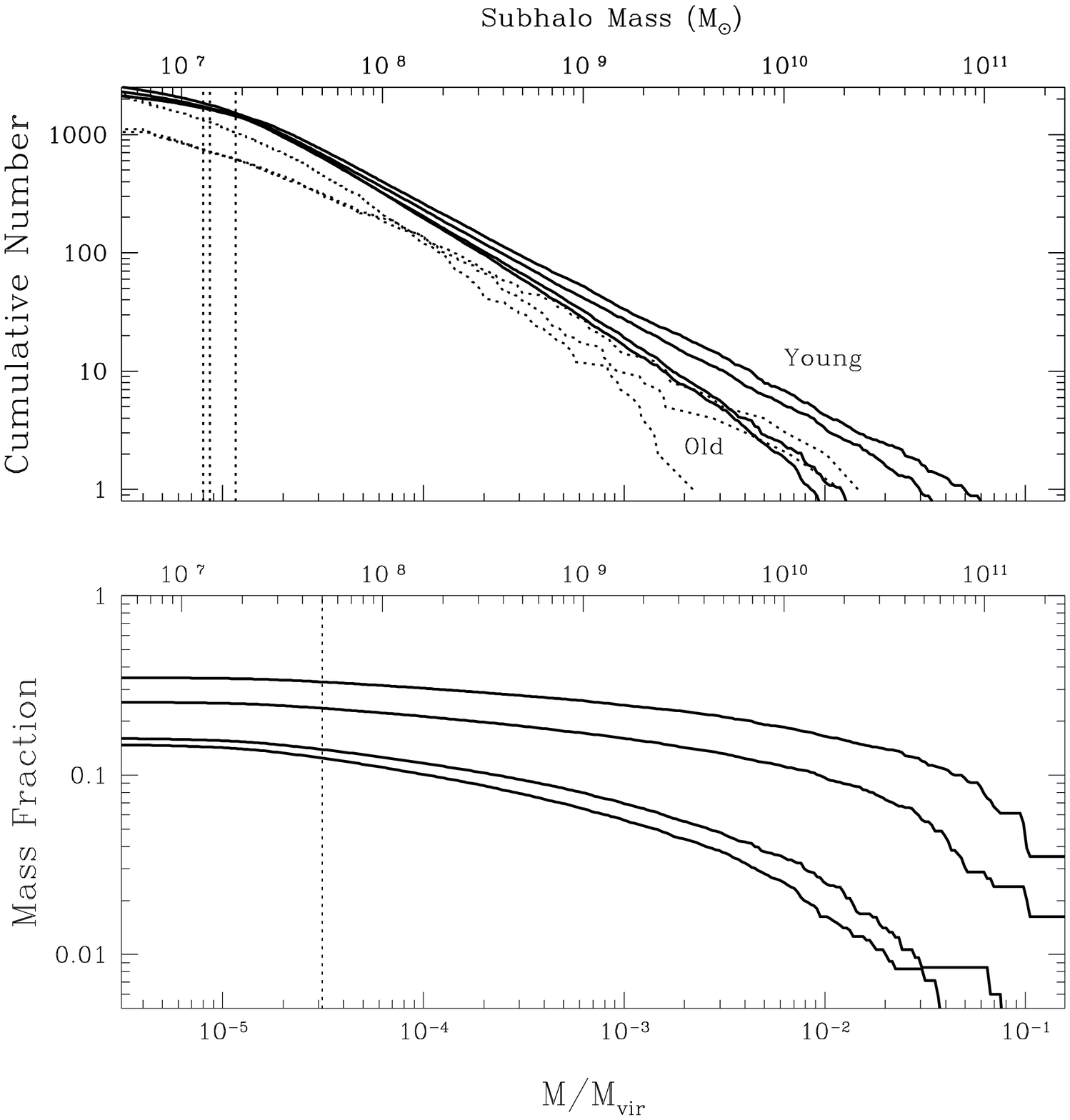}
\caption{a: ({\sl left}) The cumulative mass functions (top panels)
and cumulative mass fraction in substructure (bottom panels) in the 
highest resolution simulation (Ghigna et al.~2000; dashed lines),
compared with the average in the SA model (solid lines).
b: ({\sl right}) The cumulative mass function and cumulative mass fraction
for four sets of SA trees of increasing dynamical age.}
\end{figure}
%
\section{The Satellite-substructure Connection}
\begin{figure}
\plotfiddle{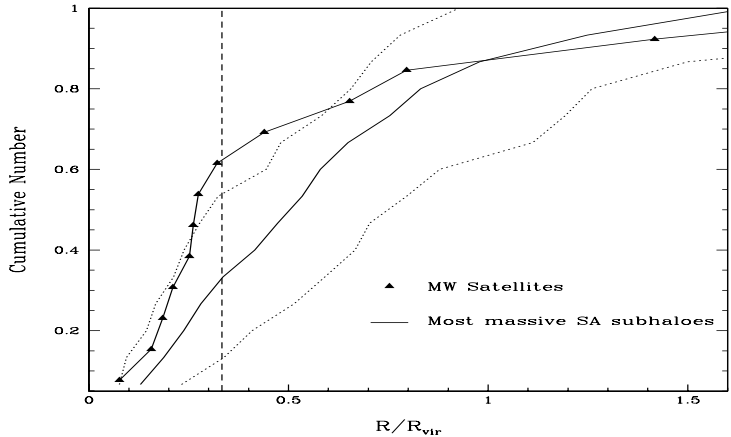}{5.5cm}{0}{50}{30}{-150}{-50}
\caption{The cumulative radial distribution of Milky Way satellites 
(triangles), compared with the average for the dozen most massive 
systems in a large set of SA halos (solid line, with the dashed lines 
indicating the 10\% and 90\% contours of the distribution).
}
\end{figure}
Many explanations have been proposed to account for
the apparent discrepancy between the large numbers of subhalos
predicted in CDM models and the few luminous satellites
seen around galaxies like the Milky Way. Two generic solutions 
are that early photoionization heated gas in low-mass halos,
halting star formation in all but the oldest of these systems 
(e.g.\ Bullock et al.\ 2000; Benson et al.\ 2002),
or alternatively that strong feedback suppresses star formation in all
low-mass halos, and that observed dwarf galaxies actually reside in
the central parts of much larger CDM subhalos (Stoehr et al.\ 2003, 
Hayashi et al.\ 2003). To distinguish between these
models, we can ask more generally: are local satellites associated with 
the most massive
subhalos, or with the oldest subhalos? Fig.\,3 shows the cumulative
radial distribution of satellites around the Milky Way, (assuming 
a virial radius of 310 kpc) compared with the average distribution
of the dozen most massive subhalos found in each of several hundred 
merger trees. The satellites are clearly more clustered
than the most massive subhalos. Two-thirds of them are within 
roughly 100 kpc of the Milky Way, for instance, whereas fewer than 
1 tree in 100 shows this degree of clustering for its most massive 
substructure. More detailed analysis (Taylor, Babul \& Silk, in preparation) 
strengthens the conclusion that the luminous satellites of the Milky Way must 
correspond to the oldest substructures in its halo, rather than the most 
massive ones.
%

\end{document}